\documentclass[12pt,a4paper,final]{iopart}
\usepackage{iopams,graphicx}
\usepackage[breaklinks=true,colorlinks=true,linkcolor=blue,urlcolor=blue,citecolor=blue]{hyperref}

\begin{document}

\title[]{Bootstrap percolation on spatial networks}

\author{Jian Gao$^1$, Tao Zhou$^{1,2}$ and Yanqing Hu$^3$}
\address{$^1$ Web Sciences Center, University of Electronic Science and Technology of China, Chengdu 610054, People's Republic of China}
\address{$^2$ Big Data Research Center, University of Electronic Science and Technology of China, Chengdu 610054, People's Republic of China}
\address{$^3$ School of Mathematics, Southwest Jiaotong University, Chengdu 610031, People's Republic of China}
\ead{yanqing.hu.sc@gmail.com}

\begin{abstract}
We numerically study bootstrap percolation on Kleinberg's spatial networks, in which the probability density function of a node to have a long-range link at distance $r$ scales as $P(r)\sim r^{\alpha}$. Setting the ratio of the size of the giant active component to the network size as the order parameter, we find a critical exponent $\alpha_{c}=-1$, above which a hybrid phase transition is observed, with both the first-order and second-order critical points being constant. When $\alpha<\alpha_{c}$, the second-order critical point increases as the decreasing of $\alpha$, and there is either absent of the first-order phase transition or with a decreasing first-order critical point as the decreasing of $\alpha$, depending on other parameters. Our results expand the current understanding on the spreading of information and the adoption of behaviors on spatial social networks.
\end{abstract}

\pacs {
       89.75.Hc 
       64.60.ah 
       05.70.Fh 
      }

\maketitle

\section{Introduction}
    Bootstrap percolation was originally introduced by Chalupa, Leath and Reich \cite{chalupa1979bootstrap} in the context of magnetic disordered systems in 1979. Since then, it has been studied extensively by physicists and sociologists, mainly due to its connections with various physical models and a variety of applications such as neuronal activity \cite{Goltsev2010Stochastic} and jamming transitions \cite{Toninelli2006Jamming}. Bootstrap percolation can be essentially considered as an activation process on networks: (\expandafter{\romannumeral1}) Nodes are either active or inactive; (\expandafter{\romannumeral2}) Once activated, a node remains active forever; (\expandafter{\romannumeral3}) Initially, each node is in an active state with a given probability $p$; (\expandafter{\romannumeral4}) Subsequently, inactive nodes become active if they have at least $k$ active neighbors; (\expandafter{\romannumeral5}) Nodes are activated in an iterative manner according to the condition in (\expandafter{\romannumeral4}), until no more nodes can be activated. This process has been investigated on different kinds of networks including lattices \cite{Kogut1981Bootstrap,Holroyd2003Sharp,ziff2009explosive,raey2012a,sausset2010bootstrap}, trees \cite{Balogh2006Bootstrap,Marek2009Metastable,Bela2014Bootstrap}, random networks \cite{baxter2010bootstrap,balogh2007bootstrap,janson2012bootstrap,amini2014bootstrap,chong2014bootstrap}, and so on.

    Bootstrap percolation has found applications in modeling the spreading of information \cite{shrestha2013message}, the propagation of infection \cite{Bizhani2012Discontinuous}, the adoption of new products and social behaviors \cite{bradonjic2013bootstrap,ree2012effects,Decelle2011Asymptotic,Gleeson2008Cascades,centola2010spread} such as trends, fads, political opinions, belief, rumors, innovations and financial decisions. For instance, one may decide to buy a product when recommended by more than $k$ users and trust a message when told by at least $k$ neighbors; cf. the well-known rule, ``What I tell you three times is true'' \cite{carroll1876hunting}. In this way, the process leads initially localized effects propagating throughout the whole network. Moreover, a broad range of generalized formulations of bootstrap percolation on social networks are investigated, such as Watts' model of opinions \cite{watts2002simple}, in which $k$ is replaced by a certain fraction of the neighbors, and disease transmission models with different degrees of severity of infection \cite{ball2005,Ball2009An}.

    Real networks are often embedded in space \cite{barthelemy2011spatial} and social networks are no exception. Previous empirical studies on online social networks \cite{liben2005geographic,goldenberg2009distance}, email networks \cite{Adamic2005How} and mobile phone communication networks \cite{lambiotte2008geographical} have confirmed a spatial scaling law, namely, the probability density function (PDF) of an individual to have a friend at distance $r$ scales as $P(r)\sim r^{\alpha}$, $\alpha\approx-1$ \cite{hu2011possible}. In fact, prior to these empirical observations, Kleinberg \cite{kleinberg2000navigation,kleinberg2000small} has proposed a spatial network model by adding long-range links to a 2-dimensional lattice, and he has proved that when $P(r)\sim r^{-1}$, the structure is optimal for information navigation. Recently, Hu \textit{et al.} \cite{hu2011possible} suggested the optimization of information collection as a possible explanation for the origin of this spatial scaling law. Although extensive studies have shown that the spatial organization can change the dimension, which dominates many important physical properties of networks \cite{Emmerich2013Complex,kosmidis2008structural,shekhtman2014robustness,huang2014navigation,li2011percolation,de2014percolation,daqing2011dimension,Xie2007Geographical,Han2011}, how it influences the spreading process on social networks under the framework of bootstrap percolation remains unclear.

    In this paper, we numerically study bootstrap percolation on Kleinberg's network, which is a typical artificial social network, being well-accepted by academic society. Setting the size of the giant active component as the order parameter, we find the distribution of long-range links' lengths can change the order of phase transition. In particular, a critical exponent is found to be $\alpha_{c} = -1$, above which a hybrid phase transition (mixed of first order and second order) is observed. Surprisingly, we find both the first-order and the second-order critical points are constant when $\alpha\ge-1$, insensitive to $k$ and other parameters. When $\alpha<-1$, the second-order critical point increases as the decreasing of $\alpha$. Meanwhile, as the decreasing of $\alpha$, there is either a decreasing first-order critical point or the absence of the first-order phase transition, which depends on $k$ and other parameters. These results indicate that the spatial scaling exponent $\alpha=-1$, observed in real social networks, may be resulted from some deep-going principles in addition to the optimization of navigation and information collection, which is not yet fully understood now.

\section{Model}

\begin{figure}[top]
\begin{center}
\includegraphics[width=9.5cm]{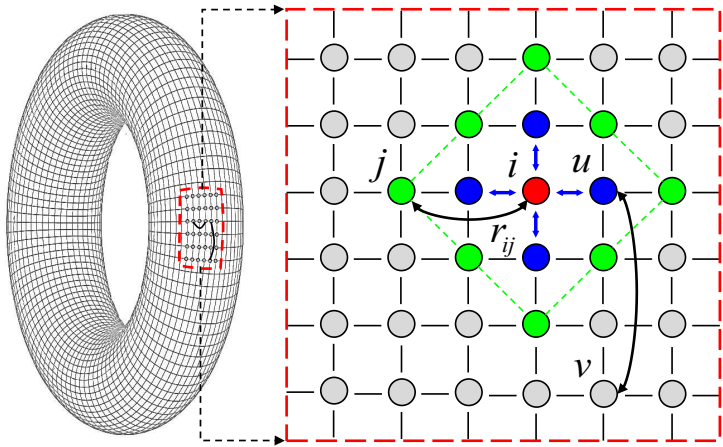}
\end{center}
\caption{An illustration of a Kleinberg's network constrained on a 2-dimensional periodic lattice. Each node has four short-range links (colored blue for node $i$) and one long-range link (colored black). The probability density function of a node to have a long-range link at Manhattan distance $r$ scales as $P(r)\propto r^{\alpha}$. For the target node $i$ (colored red), when $r=2$, there are eight candidate nodes (colored green), from which we can choose an uncoupled node $j$ to make a connection. For another target node $u$, we can choose to connect it with $v$ when $r=3$.}
\label{Fig1:lattice}
\end{figure}

    Kleinberg model \cite{kleinberg2000navigation,kleinberg2000small} is a typical spatial network model, which has been well justified by empirical data \cite{liben2005geographic,goldenberg2009distance,Adamic2005How,lambiotte2008geographical,hu2011possible}. In this paper, Kleinberg's network is constrained on a 2-dimensional periodic lattice consisting of $N=L\times L$ nodes. In addition to its initially connected four nearest neighbors, each node $i$ has a random long-range link to a node $j$ with probability $Q_{i}(r_{ij})\sim r_{ij}^{\alpha -1}$, where $\alpha$ is a tunable exponent and $r_{ij}$ denotes the Manhattan distance, which quantifies the length of the shortest path between node $i$ and node $j$, following strictly the horizontal or vertical links in lattices. Since the number of nodes at distance $r$ to a given node is proportional to $r^{d-1}$ in a $d$-dimensional lattice, the probability $Q(r_{ij})$ can be mapped to a probability density function, $P(r)\sim r^{d-1}Q(r) = r^{d-1}r^{\alpha -1} = r^{\alpha+d-2}$. In the present 2-dimensional case, the probability density function scales as $P(r)\sim r^{\alpha}$. An illustration of a 2-dimensional Kleinberg's network can be found in figure \ref{Fig1:lattice}.

    In order to numerically implement the spatial scaling law $-1$, we add long-range links to a 2-dimensional periodic lattice in a smart way as follows. First, a random length $r$ between 2 and $L$ is generated with probability $P(r)\sim r^{\alpha}$, which ensures the scaling in advance. Second, random segmentations of length $r$ to $\Delta x$ and $\Delta y$ with the only constraint that $|\Delta x| + |\Delta y| = r$ are done to determine candidate nodes, where $\Delta x$ and $\Delta y$ are both integers. Namely, for an uncoupled node $i$ with coordinates ($x$, $y$), named target node, all candidate nodes are these with coordinates ($x+\Delta x$, $y+\Delta y$) such that $|\Delta x| + |\Delta y| = r$. The above procedure ensures all candidate nodes at distance $r$ from the target node are uniformly distributed. Hence, we can randomly choose an uncoupled candidate node (i.e., a node withour any long-range link) to make an undirected link to target node $i$. In exceptional cases that all candidate nodes have been coupled, we randomly choose an uncoupled node from the whole network to accomplish the linking. We repeat such procedure for the rest uncoupled nodes until each node of the network has one undirected long-range link such that the degree of each node is exactly 5.

\section{Results}

    We focus on the following three indicators: (\expandafter{\romannumeral1}) The relative size of the giant active component ($S_{gc}$) at the equilibrium, i.e., the probability that an randomly selected node belongs to the giant active component; (\expandafter{\romannumeral2}) The number of iterations ($NOI$) to reach the equilibrium, which is usually used to determine the critical points for the first-order phase transition \cite{parshani2011critical,liu2012continuous,bashan2011percolation,Liu2012Cascading}; (\expandafter{\romannumeral3}) The relative size of the second giant active component ($S_{gc2}$), which is usually used to detect the critical points for the second-order phase transition \cite{li2011percolation,Liu2012Cascading}.

\begin{figure}
\begin{center}
\includegraphics[width=10.5cm]{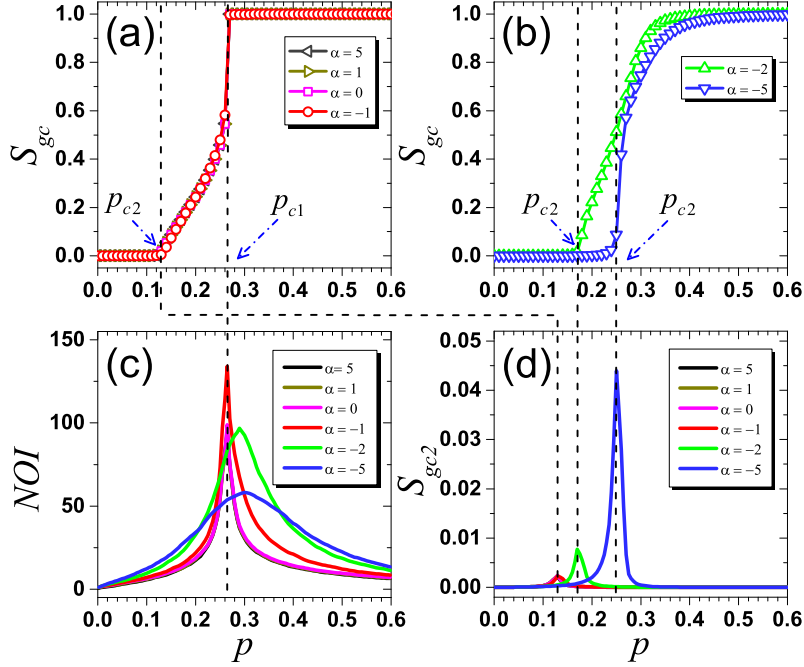}
\end{center}
\caption{$S_{gc}$ (a) and (b), $NOI$ (c) and $S_{gc2}$ (d) as a function of $p$ for different $\alpha$ after $k=3$ bootstrap percolation on Kleinberg's networks. Two different types of $S_{gc}(p)$ are observed, including a hybrid phase transition and a second-order phase transition. When $\alpha\ge-1$, $S_{gc}$ shares the same pattern and a hybrid phase transition is present. $S_{gc}$ abruptly jumps to 1 at $p_{c1}=0.263$, where $NOI$ reaches its maximum. When $\alpha<-1$, there is only a second-order phase transition with an increasing critical point as the increasing of $\alpha$, where $S_{gc2}$ reaches its maximum at different $p_{c2}$. Dash lines mark identification of critical points. Results are obtained by simulations on networks with size $400\times400$ and averaged over 1000 realizations.}
\label{Fig2:Sgc}
\end{figure}

    Figure \ref{Fig2:Sgc} shows rich phase transition phenomena when taking $S_{gc}$ as the order parameter. When $\alpha\ge-1$, $S_{gc}$ shares almost the same behavior and the system undergoes a hybrid phase transition (see figure \ref{Fig2:Sgc}(a)). We can see that $S_{gc}$ has a continue increasing at $p_{c2}=0.134$ (the second-order critical point), where the second-order phase transition is present. In contrast, $S_{gc}$ has a discontinue jump directly from about 0.60 to exact 1 at $p_{c1}=0.263$ (the first-order critical point), where the first-order phase transition occurs. To our surprise, we find that these two critical points are constant when $\alpha\ge-1$, as indicated by the four overlapping $S_{gc}-p$ curves in figure \ref{Fig2:Sgc}(a). When $\alpha<-1$, there is only a second-order phase transition with an increasing $p_{c2}$ as the decreasing of $\alpha$ (see figure \ref{Fig2:Sgc}(b)). Specifically, the second-order critical point is $p_{c2}=0.176$ when $\alpha=-2$ and $p_{c2}=0.256$ when $\alpha=-5$. Although $S_{gc}$ goes up sharper after $p$ exceeds $p_{c2}$ as the decreasing of $\alpha$, simulations justify that the curve of $S_{gc}$ is continuous. That is to say, the type of phase transition is fundamentally second-order when $\alpha<-1$.

    Finding critical points via simulations is always a difficult task that requires high precision. When $\alpha\ge-1$, where the hybrid phase transition is present, we are able to determine $p_{c1}$ by calculating the number of iterations ($NOI$) in the cascading process, since $NOI$ sharply increases when $p$ approaching $p_{c1}$ for the first-order phase transitions \cite{parshani2011critical,liu2012continuous,bashan2011percolation,Liu2012Cascading}. Accordingly, $p_{c1}$ is calculated by plotting $NOI$ as a function of $p$ (see figure \ref{Fig2:Sgc}(c)). We can see that $NOI$ reaches its maximum at the same $p$ when $\alpha\ge-1$, which is the evidence that $p_{c1}=0.263$ is constant. Analogously, by plotting the size of the second largest giant active component $S_{gc2}$ as a function of $p$, we can precisely identify $p_{c2}$ \cite{li2011percolation,Liu2012Cascading}, at which $S_{gc2}$ reaches its maximum (see figure \ref{Fig2:Sgc}(d)). It is found that $p_{c2}$ increases as $\alpha$ decreases, as $p_{c2}=0.134$ ($\alpha\ge-1$), 0.176 ($\alpha=-2$) and 0.256 ($\alpha=-5$).

\begin{figure}
\begin{center}
\includegraphics[width=10.5cm]{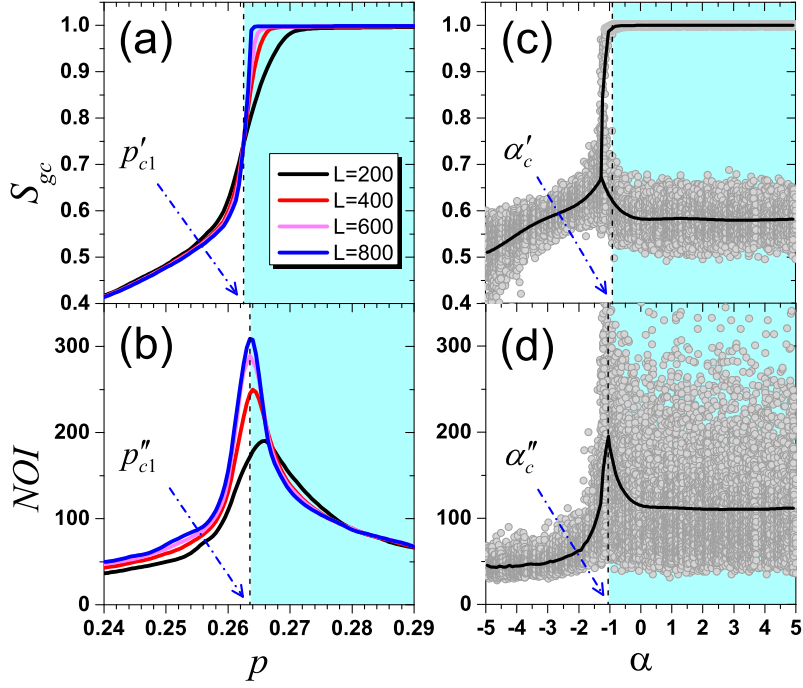}
\end{center}
\caption{Cross-validation of $p_{c1}$ and $\alpha_c$. (a) and (b) are $S_{gc}$ and $NOI$ in the case of $\alpha=-1$ under different network size $L\times L$, respectively. There is an intersection point at $p'_{c1}\approx0.2625$, while $NOI$ reaches its maximum at $p''_{c1}\approx0.2635$ in the case of $L=800$. Thus, $p_{c1}$ is identified as the average $0.263$. (c) and (d) are $S_{gc}$ and $NOI$ at $p=0.263$ under different exponent $\alpha$, respectively. For $\alpha>\alpha'_{c}\approx-0.95$, $S_{gc}$ has two phases, while $NOI$ reaches its maximum at $\alpha''_{c}\approx-1.05$. Thus, $\alpha_{c}$ is identified as the average $-1$. In figure \ref{Fig3:Points}(c) and \ref{Fig3:Points}(d), dark curves respectively represent average values of $S_{gc}$ and $NOI$, obtained from $10^4$ realizations, and each data point stands for one realization.}
\label{Fig3:Points}
\end{figure}

    Although to justify the first-order phase transition and to determine the critical exponent $\alpha_{c}$ by simulations in a discrete system are not easy, we solve this problem by a cross-validation on $p_{c1}$ and $\alpha_{c}$. Firstly, we fix $\alpha=-1$ to determine $p_{c1}$. On the one hand, there is an intersection for curves of $S_{gc}$ at $p'_{c1}\approx0.2625$ under different network sizes (see figure \ref{Fig3:Points}(a)), which can be considered as the critical point according to the finite scale analysis \cite{Liu2012Cascading,buldyrev2010catastrophic}. On the other hand, the corresponding $NOI$ reaches its maximum at $p''_{c1}\approx0.2635$ when $L=800$ (see figure \ref{Fig3:Points}(b)). Combining these two evidences, a more accurate first-order critical point is identified as $p_{c1} = (p'_{c1}+p''_{c1})/2 = 0.263$. Conversely, we fix $p=0.263$ to determine $\alpha_{c}$. From figure \ref{Fig3:Points}(c) we can see that $S_{gc}$ has two phases: exact 1 or around 0.58 when $\alpha\ge \alpha'_{c}\approx -0.95$. There is a strong evidence that $S_{gc}$ undergoes a first-order phase transition. If the increasing of $S_{gc}$ is continuous, it is impossible to observe such gap between the two phases. From figure \ref{Fig3:Points}(d), we can see that the corresponding averaging $NOI$ reaches its maximum at $\alpha=\alpha''_{c}\approx-1.05$. Combining these two evidences, we precisely identify the critical exponent as $\alpha_{c}=(\alpha'_c+\alpha''_c)/2 = -1$. In addition, the critical point for $\alpha\ge \alpha_c$ should be constant, otherwise we cannot observe the separation of two phases in figure 3(c) for a fixed value $p=0.263$.

\begin{figure}
\begin{center}
\includegraphics[width=10cm]{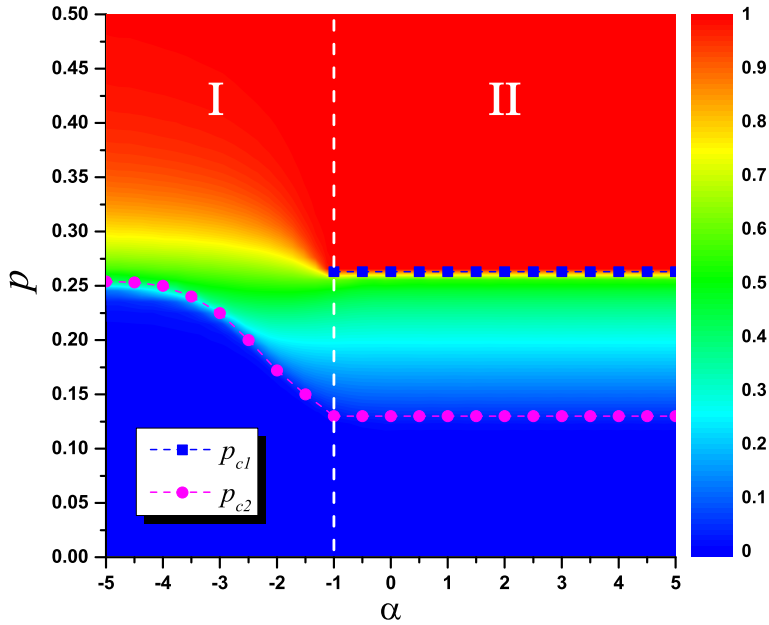}
\end{center}
\caption{Phase diagram of $k=3$ bootstrap percolation in the $p-\alpha$ plane for Kleinberg's network. The color marks $S_{gc}$. The dashed lines with solid blue squares and red circles represent the first-order critical point $p_{c1}$ and the second-order critical point $p_{c2}$, respectively. Separated by the vertical dashed line $\alpha=-1$, hybrid phase transition is observed in the right region \uppercase\expandafter{\romannumeral2} with two constant critical points $p_{c1}=0.263$ and $p_{c2}=0.134$. In the left region \uppercase\expandafter{\romannumeral1}, only second-order phase transition is present with an increasing critical point up to $p_{c2}^{\infty}=0.259$, obtained in the case of $\alpha=-\infty$, as $\alpha$ decreases. Results are averaged over 1000 realizations with fixed network size $400\times 400$.}
\label{Fig4:Phase}
\end{figure}

    A representative phase diagram for $S_{gc}$ in the $p-\alpha$ plane is shown in figure \ref{Fig4:Phase}. We find that the varying of $\alpha$, which dominates the distribution of long-range links' lengths, can change the order of phase transition. Two kinds of phase transitions are observed, consisting of the second-order phase transition and the hybrid phase transition. Overall, $\alpha_{c}=1$ is confirmed to be the critical exponent, above which the hybrid phase transition (the region labeled II) is present. Once again, we confirm that both the first-order and the second-order critical points for hybrid phase transition are constant as $p_{c1}=0.263$ and $p_{c2}=0.134$ when $\alpha\ge-1$. In the first-order phase transition region, $S_{gc}(p)$ curves are overlapped, suggesting that the properties of bootstrap percolation on Kleinberg's spatial networks have no difference if $\alpha\ge-1$. When $\alpha<-1$, the first-order phase transition is absent, leaving $S_{gc}$ undergoes only a second-order phase transition (the region labeled I) with an increasing critical point as the decreasing of $\alpha$. The maximum of $p_{c2}$ is about 0.259 when $\alpha=-\infty$, where all long-range links' lengths are 2.

    We have tested our findings for different threshold parameters $k$ and the number of long-range links associated with a node $k_{l}$. Some significant observations are robust: (i) the existence of critical point $\alpha_c=1$, above which a hybrid phase transition occurs with two constant critical points, whose values depend on $k$ and $k_l$, and (ii) the existence of a second-order phase transition when $\alpha<-1$, with an increasing critical point $p_{c2}$ as the decreasing of $\alpha$. In contrast, in some parameter spaces, the first-order transition could be observed even when $\alpha<-1$, but with a decreasing critical exponent $p_{c1}$ as the decreasing of $\alpha$. The first-order transition may disappear at a certain $\alpha$ smaller than -1. Phase diagrams of two representative examples, $(k,k_l)=(3,5)$ and $(k,k_l)=(4,7)$, are presented in figure 5.

\begin{figure}[t]
\begin{center}
    \includegraphics[height=6cm, width=6cm]{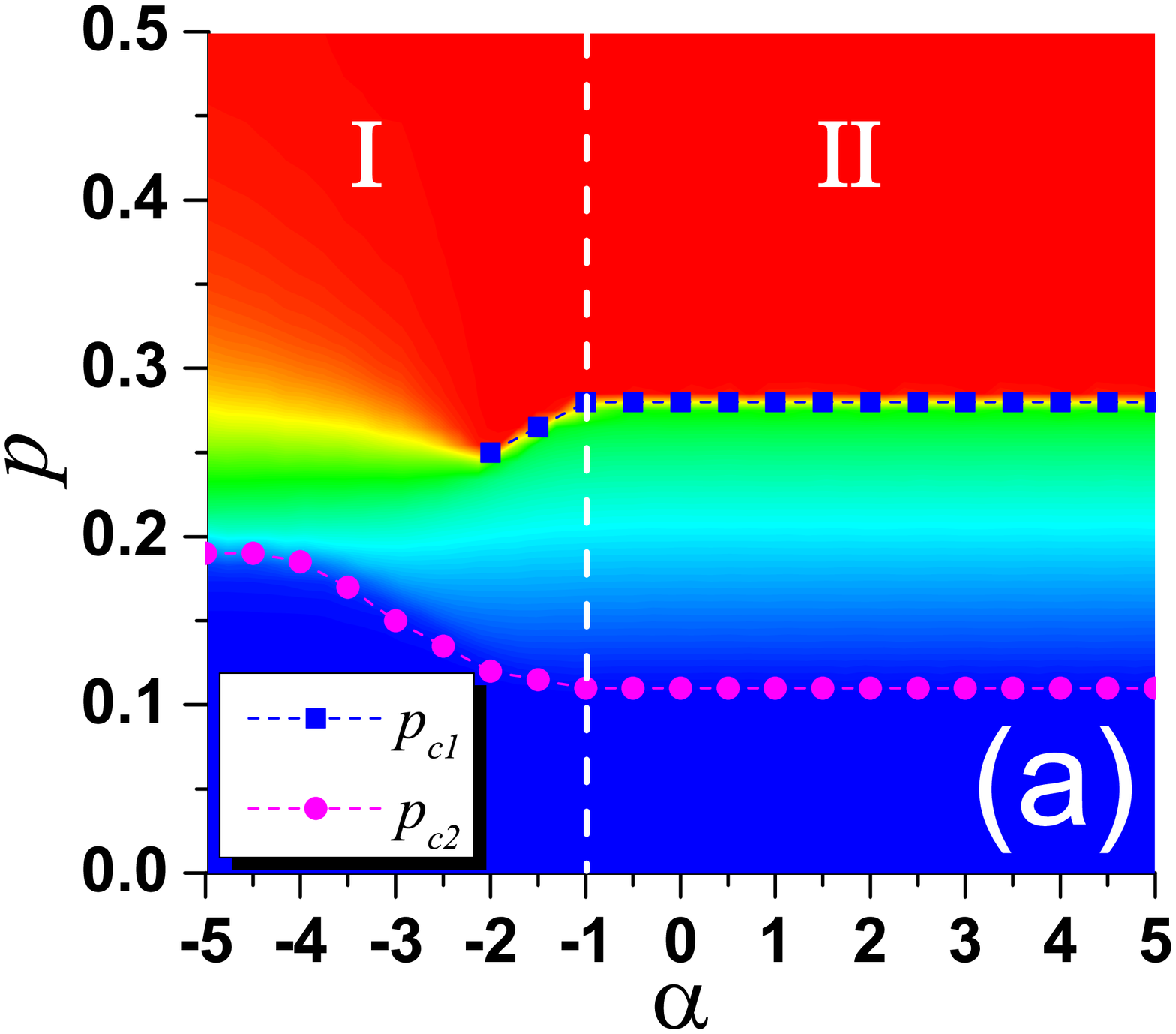}
    \includegraphics[height=6cm, width=6.3cm]{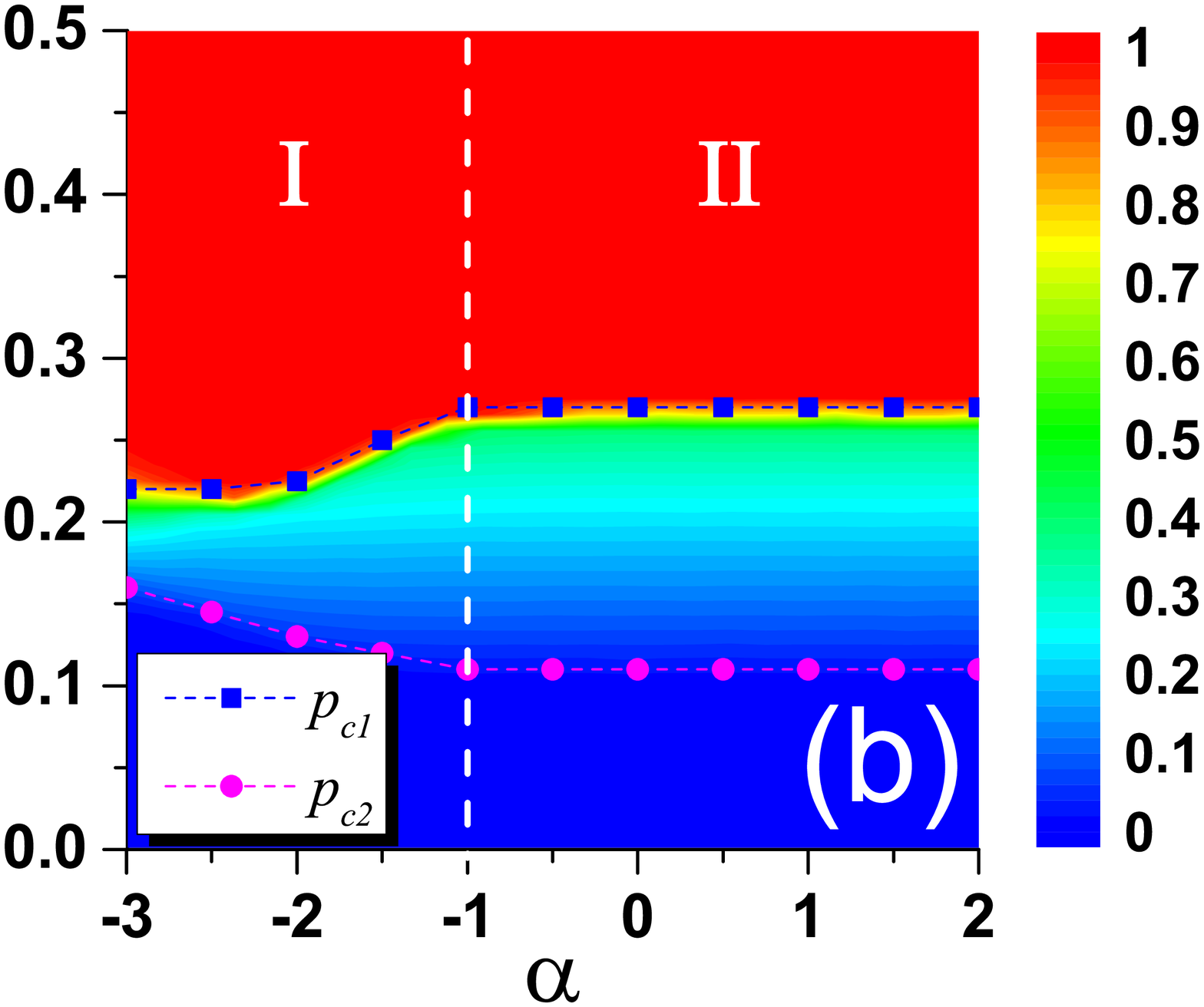}
\end{center}
\caption{Universality of the critical exponent $\alpha_c=-1$, where the color marks $S_{gc}$. (a) Phase diagram of $k=3$ bootstrap percolation in the $p-\alpha$ plane for networks where each node has $k_{l}=5$ long-range links. Separated by the vertical dashed line $\alpha=-1$, hybrid phase transition is observed in the right region \uppercase\expandafter{\romannumeral2} with two constant critical points, $p_{c1}=0.278$ and $p_{c2}=0.111$. (b) Phase diagram in the case of $k=4$ and $k_{l}=7$. The critical value is also $\alpha_c=-1$, above which a hybrid phase transition with two constant critical points, $p_{c1}=0.271$ and $p_{c2}=0.112$, is observed. Results are averaged over 1000 realizations with fixed network size $400\times400$.}
\label{Fig5:kl5-kl7}
\end{figure}

\section{Conclusions and Discussion}
    
    We have studied bootstrap percolation on spatial networks, where the distribution patterns of long-range links are found to be able to affect the order of phase transition. In particular, we find a robust critical point $\alpha_c=-1$, above which the $S_{gc}(p)$ curves are almost the same and the two critical points are constant. Such result indicates that the topological properties of Kleinberg's networks are close to each other when $\alpha>-1$ in 2-dimensional space. In fact, the critical point $\alpha=-1$ has been empirically observed in many real networks \cite{liben2005geographic,goldenberg2009distance,Adamic2005How,lambiotte2008geographical}, which may be resulted from complex self-organizing processes toward optimal structures for information collection \cite{hu2011possible} and/or navigation \cite{kleinberg2000navigation}. Since the cascading processes of Kleinberg's networks show almost the same features when $\alpha\ge-1$, this critical point $\alpha_c=-1$ is indeed corresponding to the structure with the smallest average geographical length of links, which can exhibit as effective spreading of information as networks with even longer shortcut links. This is to some extent relevant to the principle of least effort in human behavior \cite{Zipf1949}. 
    
    Our results are also relevant to the control of information spreading. For example, when $\alpha\ge\-1$, if we would like to make as more as possible people to know the information, the optimal choice of the fraction of initially informed people should be $p^*=0.263$, since larger initially informed population provides no more benefit (as shown in figure 2(a)) but requires higher cost. Furthermore, we expect to verify our findings in an analytical way based on Kleinberg's networks, spatial embedded random networks \cite{chong2014bootstrap} and multiplex networks \cite{Baxter2014Weak}. Besides, to assign each node one long-range link is a high-cost strategy when generating spatial networks, we leave individualized number of long-range links associated with each node and partial spatial embedding as future works.

\section*{Acknowledgments}
    We acknowledge Jun Wang and Panhua Huang for useful discussions. This work is partially supported by National Natural Science Foundation of China under Grants Nos. 61203156 and 11222543. J.G. acknowledges support from Tang Lixin Education Development Foundation by UESTC. T.Z. acknowledges the Program for New Century Excellent Talents in University under Grant No. NCET-11-0070, and Special Project of Sichuan Youth Science and Technology Innovation Research Team under Grant No. 2013TD0006.


\end{document}